\documentstyle[12pt]{article}

\def\be{\begin{equation}}
\def\ee{\end{equation}}
\def\bea{\begin{eqnarray}}
\def\eea{\end{eqnarray}}

\begin{document}
\title{Non-Abelian Duality and Canonical Transformations}
\author{\small
Y. Lozano,
\thanks{\tt yolanda@puhep1.princeton.edu
} \\
\small\it Joseph Henry Laboratories,\\
\small\it Princeton University,\\
\small\it Princeton, New Jersey 08544, USA}
\date{}
\maketitle
\setcounter{page}{0} \pagestyle{empty}
\thispagestyle{empty}
\vskip 1.0in
\begin{abstract}
We construct explicit canonical transformations producing non-abelian
duals in principal chiral models with arbitrary group $G$. Some
comments concerning the extension to more general $\sigma$-models,
like WZW models, are given.
\end{abstract}
\vfill
\begin{flushleft}
PUPT-1532\\
hep-th/9503045\\
March 1995
\end{flushleft}
\newpage\pagestyle{plain}

\def\theequation{\thesection . \arabic{equation}}

\section{Introduction}
\setcounter{equation}{0}

Recently some progress has been made in the understanding of target
space duality  as a symmetry of string theory in general backgrounds. Some
general review references are \cite{reviews}. These
developments started with the work of Buscher \cite{buscher} where the
dual theory of an arbitrary $\sigma$-model with an abelian isometry was
constructed. This was elaborated further in \cite{rv}, where an alternative
method of constructing the dual was presented. This method has the advantage
that it can be easily generalized to  $\sigma$-models with non-abelian
isometries, as was first done in \cite{quevedo}.
Basically it consists in gauging the isometry with some gauge fields
constrained to be flat by means of some Lagrange multipliers. Integrating
out these Lagrange multipliers and fixing the gauge the initial model is
recovered. If instead one integrates first the gauge fields and then fixes
the gauge the dual model, in which the Lagrange multipliers appear as
new coordinates, appears. Further works along these lines are
for instance \cite{mogollon,nosotros1} for the
abelian case and
\cite{nosotros2,noab} for the
non-abelian one.

For abelian duality it was shown in \cite{nosotros3} that there
exists a much simpler and straightforward way of reaching the dual theory,
and this is in terms of a canonical transformation\footnote{Canonical
transformations were first used in the context of duality
symmetries in \cite{venezia} for constant or only time-dependent
backgrounds.}.
The explicit canonical
transformation yielding the dual theory was constructed and shown to
reproduce Buscher's formulas. In this way duality can be understood as a
subgroup of the whole group of canonical transformations in the phase
space of the theory. A key point in the construction of the canonical
transformation is the fact that there exists a system of coordinates
adapted to the isometry, i.e. such that only its derivatives
appear in the Lagrangian. This turns out to be crucial in order to
obtain a local dual Hamiltonian since it allows to express its
dependence on the dual
variables and conjugate canonical momenta as local functions of the
initial variables and momenta. Plenty of useful information about the dual
model can be obtained very easily within this approach, as shown in
\cite{nosotros3}. For instance, the extension to arbitrary genus surfaces
of the proof of duality given
in \cite{buscher,rv} for spherical world-sheets and the
multivaluedness and periods of the dual variables can be
worked out by just considering
the implementation of the canonical transformation in the path
integral. Also, in the particular case of WZW-models
\cite{WZW} it becomes rather
simple to prove that the full duality group is given by
${\rm Aut}(G)_L\times {\rm Aut}(G)_R$, where $L,R$ refer to the left and
right currents on the model with group $G$, and ${\rm Aut}(G)$ are the
automorphisms of $G$, both inner and outer. It is clarified that this is
possible because the currents are chirally conserved,
this being
the reason why the
transformation leads to a local expression for the dual currents. If this
is not the case those currents not commuting with the one used to
perform duality become non-local in the dual theory and the symmetry to
which they are associated is no longer manifest.
This clarifies the point of the apparent loss of symmetries after duality
transformations.

In view of the simplicity of this approach it would be very interesting to also
have a full understanding of non-abelian duality in terms of canonical
transformations. As pointed out in \cite{nosotros3} this does not
seem obvious to do, at least as a generalization of the abelian
case,
basically because a system of coordinates adapted
to the whole set of non-commuting isometries does not exist, and as we
mentioned above this is crucial in order to obtain a local
dual Hamiltonian. However, in \cite{cz} the
non-abelian dual of the $SU(2)$ principal chiral model with
respect to the left action of the whole group was constructed
out of a canonical transformation\footnote{See also \cite{cz2}
for a supersymmetric extension.}.
This non-abelian model had been constructed before in the
literature \cite{nosotros1,fjft} following the standard procedure
described in \cite{quevedo}.
Although a system of coordinates adapted to the isometry does not
exist it is still possible to eliminate the explicit dependence on the
original variables owing to  some nice cancellations.

In this letter we present a generalization of this construction to
principal chiral models with arbitrary group $G$. Although at first sight
not obvious, it
can be seen that the explicit dependence of the Hamiltonian on the
initial variables disappears after the canonical transformation
is made. The generating functional is argued to be exact to all orders
in $\alpha^\prime$, as in the abelian case. These
general results open the possibility of understanding non-abelian
duality in terms of canonical transformations.
One of the nicest features of having a canonical
transformation description of non-abelian duality is that the
generalization to arbitrary Riemann surfaces of the result
in \cite{quevedo} for spherical world-sheets
and the multivaluedness and periods of the dual variables could
be worked out by just studying the implementation of the
transformation in the path integral.
As it was discussed in detail in
\cite{nosotros1} the main difficulty in trying to do this within the
ordinary approach of De la Ossa and Quevedo is that the
Hodge decomposition
and the splitting of the flat connection part of the auxiliary gauge
field do not coincide. It was just this coincidence what allowed
to work out this problem in the abelian case.
It would be interesting as well to obtain the group under which
the conserved currents transform under non-abelian duality
and the explicit (non-local) expressions for
the duals of the conserved currents of the initial theory.
Also, the generalization to WZW-models could be very useful in
determining further if the
theory is self-dual, as it seems to be from the semiclassical
calculations\footnote{The other possibility at hand is to
explicitly compute the partition function of the dual theory
\`a la Gawedzki and Kupiainen \cite{gk}.}.

\section{Non-abelian duality as a canonical transformation}
\setcounter{equation}{0}

Let us start this section by briefly reviewing the basic
features of abelian duality in the canonical transformation
approach.

Given a general $\sigma$-model with an abelian isometry represented by
translations of a $\theta$-coordinate:
\be
\label{ab1}
L=\frac12 g_{00} (\dot{\theta}^2-\theta^\prime\,^2)+(\dot{\theta}+
\theta^\prime)J_-+(\dot{\theta}-\theta^\prime)J_++V,
\ee
where:
\bea
\label{ab2}
&&J_-=\frac12 (g_{0i}+b_{0i})\partial_-x^i, \qquad
J_+=\frac12 (g_{0i}-b_{0i})\partial_+x^i \nonumber\\
&&V=\frac12 (g_{ij}+b_{ij})\partial_+x^i\partial_-x^j,
\eea
we can obtain the abelian dual with respect to this isometry by
performing the canonical transformation \cite{nosotros3}:
\be
\label{ab3}
p_\theta=-{\tilde \theta}^\prime, \qquad
p_{{\tilde \theta}}=-\theta^\prime.
\ee
The generating functional of this transformation is of type I and
it is given by:
\be
\label{ab4}
F=\frac12 \int_{D,\partial D=S^1} d{\tilde \theta}\wedge d\theta=
\frac12 \oint_{S^1}(\theta^\prime {\tilde \theta}-\theta {\tilde
\theta}^\prime)d\sigma.
\ee
{}From here it is possible to learn about the multivaluedness and periods of
the dual variables. In the path integral
the canonical transformation is implemented by \cite{ghandour}:
\be
\label{ab5}
\psi_k[{\tilde \theta}(\sigma)]=N(k)\int {\cal D}\theta (\sigma)
e^{iF[{\tilde \theta},\theta (\sigma)]}\phi_k [\theta(\sigma)],
\ee
where $\psi_k[{\tilde \theta}]$ and $\phi_k[\theta]$ are usually chosen as
eigenstates
corresponding to the same eigenvalue of the respective Hamiltonians and
$N(k)$ is a normalization factor. Since $\theta$ is periodic,
$\phi_k(\theta +a)=\phi_k (\theta)$
implies for ${\tilde \theta}$: ${\tilde \theta}(\sigma +2\pi)-
{\tilde \theta}(\sigma)=4\pi /a$, which means that ${\tilde \theta}$ must
live in the dual lattice of $\theta$.

This makes the duality transformation very simple conceptually, and
it also tells us how it can be applied to arbitrary genus Riemann
surfaces, because the state $\phi_k[\theta (\sigma)]$ could be the
state obtained from integrating the original theory on an arbitrary
Riemann surface with boundary. It is also clear that the arguments
generalize straightforwardly when we have several commuting
isometries.

\vspace*{1cm}

Let us now consider the principal chiral model defined by the Lagrangian:
\be
\label{uno}
L=-Tr(g^{-1}\partial_+ g g^{-1}\partial_-g)
\ee
where $g$ belongs to an arbitrary compact Lie group $G$. (\ref{uno}) is
invariant under $g\rightarrow h_1gh_2$, with $g,h_1,h_2 \in G$.
In \cite{nosotros1} the general form for the non-abelian dual with
respect to the action
of the whole group on the left was shown to be:
\be
\label{dos}
{\tilde L}=\partial_+\chi^i M^{-1}_{ij}
\partial_-\chi^j,
\ee
where $M^{ij}$ is defined by:
\be
\label{tres}
M^{ij}=\delta^{ij}+f^{ij}\,_k\chi^k
\ee
and $f^{ij}\,_k$ are the structure constants in a given basis
$\{T^k\}_{k=1}^{{\rm dim}G}$ for the Lie algebra $\underline{g}$
of $G$, normalized so that $Tr(T^k T^l)=-\delta^{kl}$.

(\ref{dos}) is invariant under
\be
\label{cuatro}
\chi^i\rightarrow R^i\,_j\chi^j,
\ee
with $R$ in the adjoint representation of $\underline{g}$.

Let us now carry out the canonical transformation approach.
The canonical coordinates we are going to use are some arbitrary
$\theta^{a}$, $a=1,\ldots,{\rm dim}G$, living in the group
manifold \cite{bow}.
We can define a matrix $\Omega^k_a(\theta)$ by:
\be
\label{cinco}
T^k \Omega^k_a (\theta)\equiv \frac{\partial g}{\partial
\theta^a}g^{-1}.
\ee
Then $\Omega\equiv T^k\Omega^k_a d\theta^a$ is the
Maurer-Cartan form on $G$.

In these variables $L$ reads:
\be
\label{seis}
L=\Omega^k_a \Omega^k_b \partial_+\theta^a
\partial_-\theta^b=\Omega^k_a\Omega^k_b
(\dot{\theta}^a\dot{\theta}^b-\theta^\prime\,^a \theta^\prime\,^b),
\ee
where sums over repeated indices are understood.

The canonical momenta are given by:
\be
\label{siete}
\Pi_a=\frac{\delta L}{\delta\dot{\theta}^a}=
2\Omega^k_a\Omega^k_b\dot{\theta}^b
\ee
and the Hamiltonian:
\be
\label{ocho}
H=\frac14\omega^{ak}\omega^{bk}\Pi_a\Pi_b+
\Omega^k_a\Omega^k_b \theta^\prime\,^a
\theta^\prime\,^b,
\ee
where the matrix $\omega^{ak}(\theta)$ is defined from:
\be
\label{nueve}
\omega^{ak}(\theta)\Omega^i_a(\theta)\equiv\delta^{ki}.
\ee
We can obtain the non-abelian dual of (\ref{uno}) with respect to the
left action of the whole group $G$ by performing the canonical
transformation generated by:
\be
\label{diez}
F[\chi,\theta]=-\oint_{S^1} \chi^i J^1_i(\theta)
d\sigma,\;\;\; i=1,\ldots,{\rm dim}G,
\ee
with $J^1_i(\theta)$ the spatial components of the conserved
currents associated
to the isometry $g\rightarrow hg$, $g,h\in G$, of the initial model.
Note that this is a type I generating functional, as in the
abelian case, and that in the particular case $g\in U(1)$ it
reduces to (\ref{ab4}).

This
generating functional produces the canonical transformation:
\bea
\label{once}
&&{\tilde \Pi}_i=\frac{\delta F}{\delta\chi^i}=-J^1_i(\theta) \nonumber\\
&&\Pi_a=-\frac{\delta F}{\delta\theta^a},
\eea
from $\{\theta^a,\Pi_a\}$ to $\{\chi^i, {\tilde \Pi}_i\}$.

As we are going to show the first equation in (\ref{once})
reflects the equality of the spatial
components of the conserved currents in the initial and dual theories.
The dual model is invariant under (\ref{cuatro}).
The conserved
currents associated to:
\be
\label{doce}
\delta\chi^i=f^i\,_{jk}\chi^j\omega^k,
\ee
with $\omega^k$ constant parameters, are given by:
\be
\label{trece}
{\tilde J}^{\mu}_i(\chi)=\frac{\delta {\tilde L}}{\delta
(\partial_\mu\omega^i)}.
\ee
The spatial components are then:
\be
\label{catorce}
{\tilde J}^1_i(\chi)=2 f^k\,_{ji}\chi^j
(M^{-1}_{[kl]} \dot{\chi}^l-
M^{-1}_{(kl)}\chi^\prime\,^l),
\ee
where $M^{-1}_{[kl]}=\frac12 (M^{-1}_{kl}-M^{-1}_{lk})$
and $M^{-1}_{(kl)}=\frac12 (M^{-1}_{kl}+M^{-1}_{lk})$.

On the other hand, the canonical momenta derived from (\ref{dos})
are given by:
\be
\label{quince}
{\tilde \Pi}_i=\frac{\delta {\tilde L}}{\delta\dot{\chi}^i}=2
(M^{-1}_{(ij)}\dot{\chi}^j-M^{-1}_{[ij]}
\chi^\prime\,^j).
\ee
Then it can be seen that if we work with the following dual
currents\footnote{These are the curvature-free currents \cite{cz},
i.e. such that $\epsilon_{\mu\nu}(\partial^\mu {\tilde I}^\nu_i
-\frac14 f^{ijk} {\tilde I}^\mu_j {\tilde I}^\nu_k)=0$, to be
compared to the initial $J^\mu_i$, also curvature free.} \cite{cz}:
\be
\label{dieciseis}
{\tilde I}^{\mu}_i={\tilde J}^{\mu}_i+2\epsilon^{\mu\nu}
\partial_\nu \chi^i
\ee
differing from ${\tilde J}^\mu_i$ in a total derivative term, we
have:
\be
\label{diecisiete}
{\tilde I}^1_i=-{\tilde \Pi}_i,
\ee
so that the first equation in (\ref{once}) is in fact:
\be
\label{dieciocho}
{\tilde I}^1_i(\chi)=J^1_i(\theta).
\ee
The conserved currents associated to the invariance
$\delta g=\epsilon^k T_k g$, with $\epsilon^k$ constant
parameters, of the initial theory are:
\be
J^\mu_i=\frac{\delta L}{\delta (\partial_\mu \epsilon^i)}=
2\Omega^i_a \partial^\mu \theta^a
\ee
so that (\ref{once}) reads:
\bea
\label{diecinueve}
&&{\tilde \Pi}_i=2\Omega^i_a \theta^\prime\,^a \nonumber\\
&&\Pi_a=2(\Omega^i_a \chi^\prime\,^i-f_{ijk}
\chi^i\Omega^j_b \Omega^k_a
\theta^\prime\,^b).
\eea

In order to obtain the dual Hamiltonian we have to express $H(\theta^a,
\Pi_a)$ as a function of $\{\chi^i,{\tilde \Pi}_i\}$ by means of the
relations (\ref{diecinueve}). This does not seem obvious to do since even
in the second equation in (\ref{diecinueve}) $\Pi_a$ appear as functions
of the initial $\{\theta^a\}$ variables. However, in (\ref{ocho})
only the
combinations $\omega^{ak}\Pi_a$, $\Omega^k_a \theta^\prime\,^a$ appear
and it is easy to see from (\ref{diecinueve}) that these are given by:
\bea
\label{veinte}
&&\omega^{ak}\Pi_a=2 \chi^\prime\,^k-\chi^i f_i\,^{jk}
{\tilde \Pi}_j \nonumber\\
&&\Omega^k_a \theta^\prime\,^a=\frac12 {\tilde \Pi}_k.
\eea
{}From here one can check that the time components of the conserved
currents of the initial and dual theories are also identified
under (\ref{diecinueve}).

The dual Hamiltonian reads:
\be
\label{veintiuno}
{\tilde H}=\frac14 {\tilde \Pi}^2+\chi^\prime\,^2-f^i_{jk}
{\tilde \Pi}_i \chi^\prime\,^j\chi^k+\frac14 f_i^{jk}
f_l^{mk}\chi^i\chi^l {\tilde \Pi}_j {\tilde \Pi}_m,
\ee
the corresponding Lagrangian being (\ref{dos}).

In the particular case $G=SU(2)$:
\be
\label{veintidos}
{\tilde H}_{SU(2)}=\frac14 {\tilde \Pi}^2+(\chi^\prime)^2+
\sqrt{2}
\epsilon^{ijk} \chi^\prime\,^k \chi^i {\tilde \Pi}_j+
\frac12 \chi^2 {\tilde \Pi}^2-\frac12 (\chi {\tilde \Pi})^2,
\ee
which is the expression obtained in \cite{cz} up to
normalization.

We can now make some comments concerning our derivation of
the canonical
transformation.

A simple consequence of our arguments is that they can readily
be extended to the action on the left of a subgroup $H\subset G$.
One just have to orthogonally decompose the Lie algebra of $G$,
$\underline{g}=\underline{h}\oplus\underline{k}$, where
$[\underline{h},\underline{k}]\subset \underline{k}$, and
write:
\be
L[g]=L[h]+L[l],
\ee
with $g=lh$, $h\in H$ and $dl l^{-1}\in\underline{k}$. Then the
same arguments can be followed for $L[h]$.

The key point which allowed us to find the coupling between the original and
dual theories was the fact that in the
particular case of principal chiral models the conserved currents of the dual
models
are given in terms of the canonical variables by just the
conjugate momenta themselves\footnote{This was first noticed
 in \cite{cz} for the case of an $SU(2)$ principal chiral model.}.  Then both
theories can be very easily related.
Of
course, in order to do this we have used some explicit information we had
a priori, which is the fact that the non-abelian duals of the
principal chiral models of group $G$ with respect to the action of the whole
group on the left had already been studied in the
literature \cite{nosotros1,fjft}. There is one feature
that is especial for principal chiral models. This is the fact
that as the
original theory is invariant under $g\rightarrow h_1 g h_2$, with
$g, h_1, h_2\in G$, and only the left action is used in order to
construct the dual the other isometry, commuting with the one acting
on the left, still remains, so that there are some conserved
currents in the dual model. Then it makes sense to try to couple
both theories by means of their conserved currents, especially
because the dual ones are so easily expressed as functions
of the canonical variables. This fact makes the extension of the
explicit construction we have just performed to arbitrary $\sigma$- models not
obvious, since in general the dual model
will not have conserved
currents at all after a non-abelian duality transformation is made. This is
what happens for instance in the case of WZW models, where the only
non-anomalous action that can be gauged is the vectorial
action, which yields dual models without conserved currents
(at least locally). The generalization to these models
remains then an interesting open problem which we hope to address in a future
publication.

The generating functional (\ref{diez}) is linear in the dual variables
but not in the original ones, so we expect a priori that it will
receive quantum corrections when implemented in the path integral
\cite{ghandour}:
\be
\label{dis1}
\psi_k [\chi^i(\sigma)]=N(k)\int\prod_{a=1}^{{\rm dim}G}{\cal D}\theta^a
(\sigma)e^{iF[\chi,\theta (\sigma)]}\phi_k [\theta^a (\sigma)].
\ee
This formula is the starting point to work out the multivaluedness and
periods of the dual variables. This information is not available at
present for general non-abelian duality transformations,
as we  mentioned
in the introduction. Only in the particular case of $\sigma$-models with
chiral currents and as a consequence of the Polyakov-Wiegmann property
\cite{pw} satisfied by WZW models we know the space in which
the dual variables live \cite{nosotros2}. Within the approach of the canonical
transformation
one needs to compute the quantum corrections to the generating
functional (\ref{diez}) in order to obtain this information.

It is easy to check that the following relation holds\footnote{This was
first noticed in \cite{cz} for $G=SU(2)$.}:
\be
\label{dis2}
{\tilde I}^i_\mu [\chi]\psi_k [\chi^i(\sigma)]=
N(k)\int\prod_{a=1}^{{\rm dim}G}{\cal D}\theta^a (\sigma)e^{iF[\chi,\theta]}
J^i_\mu [\theta] \phi_k[\theta^a (\sigma)],
\ee
with ${\tilde I}^i_\mu$ given by (\ref{dieciseis}). So if we choose
$\psi_k[\chi]$ and $\phi_k[\theta]$ as eigenfunctions of the respective
conserved currents with the same eigenvalue, (\ref{dis1}) holds with
$F$ given by the classical expression (\ref{diez}). This implies that
although the generating functional is not linear in $\{\theta^a\}$ it is
in fact exact to all orders in $\alpha^\prime$. However this argument
is formal and in order to further establish that
$F$ is generating a quantum
transformation one would need to take into account renormalization
effects.

\subsection*{Acknowledgements}

I would like to thank E. Alvarez, L. Alvarez-Gaum\'e, J.L.F. Barb\'on
 and C. Zachos
for useful discussions. A Fellowship from M.E.C. (Spain)
is acknowledged for partial financial support.

\end{document}